\documentclass[aps,showpacs,superscriptaddress,twocolumn,amsmath,amssymb,pre]{revtex4-1}
\usepackage{dcolumn}
\usepackage{yfonts}
\usepackage{bm}
\usepackage{amssymb}
\usepackage{graphicx}
\usepackage{color}
\usepackage{wrapfig,subfigure}
\usepackage{ulem}

\newcommand{\qS}{q_{\cal S}}

\newcommand{\ep}{\varepsilon}
\newcommand{\ka}{\varkappa}

\def\p12{p_{12}({\bf q},t)}

\begin{document}
\title{
Large rare fluctuations in systems with delayed dissipation
}
\author{M. I. Dykman}
\affiliation{Department of Physics and Astronomy,
Michigan State University, East Lansing, MI 48824}
\author{I. B. Schwartz}
\affiliation{US Naval Research Laboratory, Code 6792, Nonlinear System Dynamics Section, Plasma Physics Division, Washington, DC 20375}
\date{\today }

\begin{abstract}
We study the probability distribution and the escape rate in systems with delayed dissipation that comes from the coupling to a thermal bath. To logarithmic accuracy in the fluctuation intensity, the problem is reduced to a variational problem. It describes the most probable fluctuational paths, which are given by acausal equations due to the delay. In thermal equilibrium, the most probable path passing through a remote state has time reversal symmetry, even though one cannot uniquely define a path that starts from a state with given system coordinate and momentum.  The corrections to the distribution and the escape activation energy for small delay and small noise correlation time are obtained in the explicit form. 
\end{abstract}

\pacs{ 05.40.-a, 02.50.Ey, 85.85.+j, 81.07.Oj}
\maketitle
\section{Introduction}

Large fluctuations play an important role in many physical phenomena, an example being spontaneous switching between coexisting stable states of a system, like switching between the magnetization states in magnets, or voltage/current states in Josephson junctions, or macromolecule configurations or populations. Typically, large fluctuations are rare events on the dynamical time scale of the system. A theoretical analysis of such events goes back to Kramers \cite{Kramers1940}, who considered the rate of switching of a Brownian particle from a potential well. The problem of the switching rate and the probability distribution becomes more complicated for systems away from thermal equilibrium, as in this case the probability distribution is no longer of the Boltzmann form. A rigorous mathematical approach to the problem was developed and many results have been obtained for dynamical systems without delay driven by white Gaussian noise and for Markovian reaction and population systems, cf. \cite{Freidlin_book,Wentzell1976,Graham1984a,*Graham1989a,Dykman1980,*Dykman1994c,*Dykman2008,Day1987a,Jauslin1987,Maier1993a,*Maier1997,Touchette2009,Kamenev2011}.  More recently the problem of large rare fluctuations in white-noise driven systems with delay was addressed in the mathematical literature \cite{Mohammed2006,*Kushner2010}.

Delay naturally arises in dissipative dynamical systems. In such systems, dissipation results from the coupling to a reservoir: motion of the system causes changes in the reservoir, which in turn affect the motion. The underlying reaction of the reservoir is generically delayed. Along with the dissipative force, the reservoir exerts a random force on the system. If dissipation is delayed, the random force has a finite correlation time. These effects have been attracting much attention in the context of optomechanics and the dynamical back-action \cite{Kippenberg2008}.

In this paper we develop a formalism for studying large rare fluctuations in classical systems with delayed dissipation. Much of the analysis refers to the case of Gaussian noise, but the results can be immediately extended to non-Gaussian noise as well. An important part of the paper, that allows us to test the general formulation, is the analysis of coupling to a reservoir in thermal equilibrium, where the noise and the dissipative force are connected by the fluctuation-dissipation relation \cite{LL_statphys1}.

Central for the analysis is the idea of the optimal fluctuation. In a large rare fluctuation the system is brought from its stable state to a remote state in phase space. This requires a large deviation of the noise from its root mean square value. Different noise realizations can result in the same outcome, but they have different probability densities; for Gaussian noise, the difference is exponential. The overall probability of a large fluctuation of the system is determined by the most probable, or optimal appropriate realization of the noise. 

As a consequence, in a fluctuation to a remote point in phase space or in switching the system is most likely to move along a well-defined (optimal) trajectory, that corresponds to the optimal noise realization. Using the approach \cite{Dykman1992d} and its extensions, the narrow peak in the distribution of the trajectories has indeed been seen in simulations and in the experiments, cf. \cite{Luchinsky1997b,Vugmeister1997a,*Mannella1999,Morillo1997,Hales2000,*Ray2006,Chan2008,Forgoston2009}.

An important feature of large rare fluctuations in Markovian (no delay) systems in thermal equilibrium with a bath is  that the optimal fluctuational path is the time reversed path in the absence of noise, cf. \cite{Onsager1953,*Machlup1953}.  This can be understood from the argument that, in relaxation in the absence of noise, the energy of the system goes into the entropy of the thermal reservoir, whereas in a large fluctuation the entropy of the reservoir goes into the system energy. The minimal entropy change corresponds to a time-reversed process \cite{LL_statphys1}. In other words, the optimal trajectory for a large fluctuation corresponds to the noise-free trajectory with the inversed sign of the friction coefficient. One can view this property also as a consequence of the symmetry of transition rates in systems with detailed balance discussed for diffusive systems described by the Fokker-Planck equation by Kolmogorov \cite{Kolmogoroff1937} (optimal fluctuational paths and the path distribution were not discussed in Ref.~\onlinecite{Kolmogoroff1937}). 

We show below that the situation is more complicated if the dissipative force is delayed. We consider linear coupling to a thermal reservoir, which leads to a delayed viscous friction. The model is described in Sec.~\ref{sec:model} below. A variational problem for finding optimal fluctuational paths in systems with delayed relaxation in the presence of Gaussian noise is formulated in Sec.~\ref{sec:variational_problem}. Both the problems of the tail of the probability distribution and escape from a metastable state are considered. In Sec.~\ref{sec:thermal_noise} we show that, if the noise has thermal origin, the tail of the probability distribution remains to be of the Boltzmann form in the presence of dissipation delay. We also consider the time-reversal symmetry of the most probable trajectories. In Sec.~\ref{sec:expon_correlated_noise} the results are illustrated using an exponentially correlated thermal noise. In Sec.~\ref{sec:short_corr_time} we give explicit expressions for the logarithm of the probability distribution and the escape activation energy for the case where the correlation time of the noise and the dissipation delay are short. Sec.~\ref{sec:conclusions} contains concluding remarks.

\section{A system linearly coupled to a thermal bath}
\label{sec:model}

We start by sketching the derivation of the equation of motion with delayed dissipation for a particle coupled to a thermal bath. This problem was addressed back in the mid-60s \cite{Mori1965,Kubo1966}. In contrast to Refs.~\onlinecite{Mori1965,Kubo1966}, our analysis refers to a classical particle in a generally {\it nonparabolic} potential $U(q)$, where $q$ is the particle coordinate. The particle has a unit mass and the bath has a quasi-continuous frequency spectrum. The coupling is linear in the particle coordinate $q$. The total Hamiltonian of the system and the bath is
\begin{eqnarray}
\label{eq:full_Hamiltonian}
H&=& H_0 + H_{\rm b} + H_i; \qquad H_0=\frac{1}{2}p^2 + U(q); \nonumber\\ 
H_i&=&qh_{\rm b}.
\end{eqnarray}
Here, $p$ is the momentum of the particle; $H_{\rm b}$ is the Hamiltonian of the bath in the absence of the interaction; $h_{\rm b}$ is a function of the dynamical variables of the bath only. A familiar example is a bath that consists of harmonic oscillators, with $h_{\rm b}$ linear in the oscillator coordinates $q_k$, cf. \cite{Bogolyubov1945,Feynman1963,FeynmanQM,Ford1965},
\begin{equation}
\label{eq:Feynman_vernon}
H_{\rm b}=\frac{1}{2}\sum_k(p_k^2+\omega_k^2q_k^2), \qquad h_{\rm b}=\sum_k\ep_kq_k.
\end{equation}
The analysis below is not limited to this model; it immediately extends also to the case where the interaction $H_i$ is linear in the particle momentum $p$.

In the equation of motion of the system
\begin{equation}
\label{eq:trivial_eom}
\ddot q(t) + U'\bigl(q(t)\bigr) + h_{\rm b}(t)=0.
\end{equation}
function $h_{\rm b}(t)$ itself depends on $q(t')$ with $t'\leq t$, because the bath is perturbed by the system. In much of the paper we assume the interaction to be weak, i.e., that $h_{\rm b}$ is proportional to a small constant. Then the response of the bath to the motion of the system can be described using the  generalized susceptibility $\alpha_h$. It determines the mean value of $h_{\rm b}$ if, instead of the coupling to the considered dynamical system, the bath were driven by a time-dependent force $F(t)$, with energy $-F(t)h_{\rm b}$. In the considered case the role of $F(t)$ is played by $-q(t)$, 
\begin{equation}
\label{eq:linear_response}
h_{\rm b}(t) \approx h_{\rm b}^{(0)}(t) - \int_{-\infty}^tdt' \alpha_h(t-t')q(t'),
\end{equation}
where we assumed that the interaction was adiabatically turned on at $t\to -\infty$. In the model (\ref{eq:Feynman_vernon}) $\alpha_h(t)=\sum_k(\ep_k^2/\omega_k)\sin\omega_kt$ \cite{Bogolyubov1945,Ford1965}; Eq.~(\ref{eq:linear_response}) applies in this case for an arbitrarily strong coupling.

 In Eq.~(\ref{eq:linear_response}), $h_{\rm b}^{(0)}(t)$ is the value of $h_{\rm b}(t)$ in the absence of the interaction with the system. It is a random function of time; for example, in the model (\ref{eq:Feynman_vernon}) the randomness comes from the randomness of the amplitudes and phases of the non-interacting oscillators \cite{Bogolyubov1945}. We set $\langle h_{\rm b}^{(0)}(t)\rangle = 0$. The power spectrum of $h_{\rm b}^{(0)}$ can be written as $2{\text Re}~\Phi_h(\omega)$, where
\begin{eqnarray}
\label{eq:power_spectrum}
&&\Phi_h(\omega)=\pi^{-1}\int\nolimits_{0}^{\infty}dt \exp(i\omega t)\phi_h(t),\nonumber\\
&&\phi_h(t)=\langle h_{\rm b}^{(0)}(t)h_{\rm b}^{(0)}(0)\rangle.
\end{eqnarray}
The power spectrum is related to the susceptibility $\alpha_h$ by the fluctuation-dissipation theorem \cite{LL_statphys1}. 

If $h_{\rm b}$ is nonlinear in the dynamical variables of excitations of the bath, Eq.~(\ref{eq:linear_response}) can be obtained by calculating the first-order in $H_i$ correction to these variables and expanding $h_{\rm b}$ to the first order in this correction.  Function $\alpha_h(t-t')$ is determined by the mean value of the coefficient at $q(t')$ calculated in the absence of the system-to-bath coupling. The remaining random part is of higher order in the interaction than $h_{\rm b}^{(0)}$ and therefore is disregarded. As will be shown separately, the decoupling that leads to Eq.~(\ref{eq:linear_response}) applies if the correlation time of fluctuations of the bath is small compared to the effective relaxation time of the system, which is determined by the interaction $H_i$. For the coupling (\ref{eq:Feynman_vernon}) there is no this limitation.

Using that, from the fluctuation-dissipation theorem, $\alpha_h(t)=-\beta \dot\phi_h(t)$ for $t>0$  ($\beta=1/k_BT$), one obtains from Eqs.~(\ref{eq:full_Hamiltonian}) and (\ref{eq:linear_response}) the equation of motion of the system coupled to the bath in the form
\begin{eqnarray}
\label{eq:eom}
\ddot q(t)& +& U_h'\bigl(q(t)\bigr) + \beta \int_{0}^{\infty}dt' \phi_h(t')\dot q(t-t') = f(t), \nonumber\\
&&U_h(q) = U(q)-(\beta/2)q^2\phi_h(0).
\end{eqnarray}
Here $f(t)$ is a random force; if the only source of this force is the coupling to the bath, then $f(t)=h_{\rm b}^{(0)}(t)$. However, often in the experiment noise comes from external sources that are not in thermal equilibrium with the system, and $f(t)$ accounts for such noise.

The integral term in Eq.~(\ref{eq:eom}) describes dissipation of the system due to the coupling to a thermal bath. The dissipation is delayed. Within the linear response approximation it is linear in the velocity of the system, but depends on the velocity history. The coupling to the bath leads also to the renormalization of the potential of the system $U(q)\to U_h(q)$. We note that it is natural to count $q$ off from a minimum of the ``bare" potential $U(q)$ (a constant shift of $q$ can be incorporated into $H_{\rm b}$). Then the renormalization (\ref{eq:eom}) corresponds to softening of the potential near this minimum, since $\phi_h(0)>0$.

\subsection{Stationary states}
\label{subsec:stationary_states}

The stationary states of the system (\ref{eq:eom}) in the absence of noise are located at the extrema of $U_h(q)$. Near its extremum $q_0$ the potential $U_h$ can be linearized in $\delta q=q-q_0$ and the solution of  Eq.~(\ref{eq:eom}) can be sought in the form $\delta q(t)\propto\exp(\lambda t)$ with $\lambda$ given by equation
\begin{equation}
\label{eq:eigenvalues}
\lambda^2 + U_h''(q_0) +\pi \beta\lambda\Phi_h(i\lambda)=0.
\end{equation}

Equation (\ref{eq:eigenvalues}) is simplified if the coupling to the bath is so weak that the last term is a perturbation compared to $U_h''$. In this case, if $U_h$ has a minimum at $q_0\equiv q_a$, with $U_h''(q_a)\equiv \omega_a^2 >0$, 
\begin{equation}
\label{eq:stable_eigenvalues}
\lambda_{a\pm} \approx \pm i\tilde\omega_a -\Gamma,\qquad \Gamma=(\pi\beta/2){\text Re}~\Phi_h(\omega_a),
\end{equation}
where $\tilde\omega_a = \omega_a+(\pi\beta/2){\text Im}~\Phi_h(\omega_a)$. Equation (\ref{eq:stable_eigenvalues}) applies provided $\beta|\Phi_h(\omega_a)|\ll \omega_a$.

Since the thermal noise power spectrum $2{\rm Re}~\Phi_h(\omega)$ is non-negative, $\Gamma\geq 0$. Therefore in the presence of delay a minimum of $U_h$ still corresponds to an asymptotically stable state of the system, an attractor. Parameter $\Gamma$ is the characteristic relaxation rate of the system. Since $\Gamma \ll \omega_a$, the stable state is a focus on the phase plane $(q,p)$ and the motion near $(q_a,p_a=0)$ is underdamped. 

Generally, systems with delay have an infinite-dimensional phase space. Therefore  Eq.~(\ref{eq:eigenvalues}) should have more than two solutions. However, function $\Phi_h(\omega)$ is analytical for Im~$\omega > 0$. Then  $\Phi_h(i\lambda)$ does not diverge for Re~$\lambda>0$, and if the coupling is small, the last term in Eq.~(\ref{eq:eigenvalues}) remains small for Re~$\lambda >0$, which means that there are no solutions of Eq.~(\ref{eq:eigenvalues}) with Re~$\lambda >0$ in the weak-coupling limit. In turn, this means that a minimum of $U_h(q)$ is an attractor. 

In many cases of physical interest the correlator $\phi_h(t)$ exponentially decays for large times, $\phi_h(t)\propto\exp(-t/t_c)$ for $t \to \infty$, where $t_c$ is the correlation time of bath fluctuations. Then for a very weak coupling  Eq.~(\ref{eq:eigenvalues}) has a root Re~$\lambda \approx -1/t_c$, which describes a comparatively fast relaxation of the bath when the system is close to the attractor, $\Gamma t_c\ll 1$. We will not discuss in this paper the case of a power-law decay of correlations in the bath, which has attracted much attention in the context of quantum tunneling \cite{Leggett1995}, although some of the results apply to this case, too.

A local maximum of $U_h(q)$ is a saddle point; we will use the notation $\qS$ for this point. Within a naive perturbation theory,  near the saddle point  in the limit of small $|\Phi_h|$ the eigenvalues are
\begin{equation}
\label{eq:saddle_eigenvalues}
\lambda_{{\cal S}\pm} \approx \pm \Omega_{\cal S} -(\pi\beta/2)\Phi_h\left(\pm i\Omega_{\cal S})\right),
\end{equation}
where $\Omega_{\cal S} = |U''(\qS)|^{1/2}$. The coupling-induced renormalization of the root $\lambda_{{\cal S}+}$, that describes moving away from $\qS$, is small for small $|\Phi_h|$. However, the change of the root $\lambda_{{\cal S}-}$ can be significant even in the small-$|\Phi_h|$ limit. Indeed, if $\phi_h(t) \propto \exp(- t/t_c)$ for $t\to \infty$ and $t_c^{-1} <|U_h''(\qS)|^{1/2}$, the correction $\propto \Phi_h$ to $\lambda_{{\cal S}-}$ in Eq.~(\ref{eq:saddle_eigenvalues}) diverges: formally, $\Phi_h(\omega)$ diverges for $-{\rm Im}~\omega > 1/t_c$. Physically, the system moves too quickly for the bath to follow it. In this case Re~$\lambda_{{\cal S}-}$ approaches $ -1/t_c$ in the weak-coupling limit, i.e., an arbitrarily weak coupling to the bath leads to a finite change of the ``stable" eigenvalue near the saddle point. 

A dramatic change of the dynamics can occur also for the coupling of the form (\ref{eq:Feynman_vernon}), if the bath frequencies form a finite-width band \cite{Dhar2007}. We will not consider this case here.

Even where the coupling to the bath is weak in the sense that the decay of the system is slow compared to the decay of correlations in the bath, the relaxation rate of the system  can exceed the frequency $\omega_a$. In this case, for $\Phi_h(\omega)$ smooth near $\omega=0$, the motion near the potential minimum is overdamped, 
\begin{eqnarray}
\label{eq:overdamped_eigenvalue}
\lambda_{a+}\approx  -\omega_a^2/\pi\beta\Phi_h(0), \qquad 
\lambda_{a-} \approx -\pi\beta\Phi_h(0),
\end{eqnarray}
with $|\lambda_{a-}|\gg \omega_a \gg |\lambda_{a+} |$. Obviously, the eigenvalues $\lambda_{a\pm}$ are real and negative, indicating that the potential minimum remains an attractor. We call the corresponding parameter range the small inertia limit. Indeed, if we incorporate the particle mass $m$ into the Hamiltonian $H_0$ and define $U_h''(q_a)=m\omega_a^2$, we see that $\lambda_{a-}\propto 1/m$, whereas $\lambda_{a+}\propto m$. The root $\lambda_{a+}$ characterizes the slow motion of the particle coupled to the bath, other degrees of freedom relax much faster.

Similarly, near $\qS$ for small inertia  we have
\begin{equation}
\label{eq:overdamped_saddle}
\lambda_{{\cal S}-} \approx -\pi\beta\Phi_h(0),\qquad \lambda_{{\cal S}+} \approx |U_h''(\qS)|^2/\pi\beta\Phi_h(0)
\end{equation}
with $|\lambda_{{\cal S}-}|\gg |U_h''(\qS)| \gg |\lambda_{{\cal S}+} |$. The potential maximum remains a saddle point, with motion away from it being slow compared to the motion toward it. Approaching the saddle point is characterized, essentially, by the relaxation rate of the bath when the system is at the saddle point. Of relevance to the motion of the system in the limit of small mass is primarily the root $\lambda_{{\cal S}+}$.

If the coupling is described by the model (\ref{eq:Feynman_vernon}) and is not weak, we have not found a simple explicit expression for the eigenvalues $\lambda$. However, one can expect that the minima of $U_h(q)$ remain stable states. This is a consequence of the condition Re~$\Phi_h(\omega)>0$ for Im~$\omega=0$. Indeed, because of this condition Eq.~(\ref{eq:eigenvalues}) has no roots with Re~$\lambda = 0$ for $q_0=q_a$. Therefore, since for weak coupling Re~$\lambda_a <0$, and given that the dependence of $\lambda_a$ on the coupling strength is smooth, as Eq.~(\ref{eq:eigenvalues}) suggests,  as we increase the coupling strength the roots $\lambda_a$ will never cross the axis Re~$\lambda_a=0$. Hence, the state $(q_a,p_a)$ will remain stable, see an example in Sec.~\ref{sec:expon_correlated_noise}.

\section{Variational problem for optimal fluctuations}
\label{sec:variational_problem}

Large rare fluctuations in systems with delayed dissipation and with non-white noise can be analyzed by extending the approach developed for systems with no delay. The underlying idea is that, before a large fluctuation occurs, the system performs small-amplitude fluctuations about its initially occupied stable state. These small fluctuations persist for a long time that largely exceeds the relaxation time. When ultimately there occurs a large fluctuation to a given point $(q,p)$ in phase space, the system is most likely to move along the optimal trajectory that corresponds to the most probable appropriate realization of the noise, as outlined in the Introduction. 

If the noise $f(t)$ in Eq.~(\ref{eq:eom}) is Gaussian and stationary, with zero mean and correlation function
$\phi_f(t-t')=\langle f(t)f(t')\rangle$,
the probability density functional of noise realizations is   ${\cal P}_f[f(t)]=\exp\Bigl(-{\cal R}_f[f(t)]/D\Bigr)$ \cite{FeynmanQM}, where
\begin{eqnarray}
\label{eq:noise_distribution}
&&{\cal R}_f[f(t)]=\frac{1}{4}\int\nolimits_{-\infty}^{\infty}dt\,dt'f(t) {\cal F}(t-t') f(t'),\nonumber\\
&&\int\nolimits_{-\infty}^{\infty}dt_1{\cal F}(t-t_1)\phi_f(t_1-t')=2D\delta(t-t').
\end{eqnarray}
Here, $D$ is the characteristic noise intensity; in the case of thermal noise, $f(t) = h_{\rm b}^{(0)}(t)$ and $D=k_BT$. We assume that $D$ is small, so that on average the amplitude of fluctuations of the system about its attractor is small. Function ${\cal F}(t)/2D$ is the inverse of the noise correlator $\phi_f(t)$.

To logarithmic accuracy, the probability density of reaching a state $(q,p)$ is 
\begin{equation}
\label{eq:prob_density}
\rho(q,p)={\rm const}\times\exp\left[-R(q,p)/D\right],
\end{equation}
where $R(q,p)$ is the minimum of the functional ${\cal R}_f$ with respect to the noise trajectories that bring the system from the attractor to the state $(q,p)$. As for Gaussian-noise driven systems without delay \cite{Dykman1990},  $R(q,p)$ is given by a solution of the variational problem
\begin{eqnarray}
\label{eq:var_problem}
R(q,p)&=&\min\left\{ {\cal R}_f[f(t)] + \int\nolimits_{-\infty}^{\infty}dt\,\chi(t)\left[\ddot q(t) + U'_r\bigl(q(t)\bigr)\right.\right. \nonumber\\
&&\left.\left.+\beta\int\nolimits_{-\infty}^tdt' \phi_h(t-t')\dot q(t') - f(t)\right]\right\},
\end{eqnarray}
where the minimum is taken with respect to functions $q(t), f(t)$, and $\chi(t)$. The auxiliary function $\chi(t)$ is a Lagrange multiplier: it accounts for the interrelation between the trajectory of the system $q(t)$ and the force trajectory $f(t)$, which is given by Eq.~(\ref{eq:eom}). 

Equation (\ref{eq:var_problem}) describes the optimal fluctuational trajectories of the system and the noise. The boundary conditions for the trajectories where the system arrives from the attractor $(q_a,p_a=0)$ occupied for $t\to -\infty$ to a given $(q,p)$ at a given time $t$ (we set $t=0$)  are
\begin{widetext}
\begin{eqnarray}
\label{eq:boundary_conditions}
f(t),\chi(t)\to 0,\quad q(t)\to q_a, \quad p(t)\to p_a\quad {\rm for} \quad t\to -\infty;\nonumber\\
f(t)\to 0,\; t\to \infty; \qquad \chi(t)=0, \;t>0;  \qquad q(t=0)=q,\qquad p(t=0)=p.
\end{eqnarray}
\end{widetext}
The boundary condition for $t=0$ corresponds to the picture of optimal fluctuation to a point $(q,p)$ in which, once the system has reached this point, its further dynamics is no longer relevant. Respectively, the force should evolve for $t>0$ so as to minimize ${\cal R}_f$ independent of how its evolution affects the system. This is formally described by setting $\chi=0$ for $t>0$. Alternatively, one can set the upper limit of the integral over $t$ in Eq.~(\ref{eq:var_problem}) equal to zero, see below. Since ${\cal R}_f$ is  positive definite, its minimum is reached for $f(t)=0$, and therefore $f\to 0$ for $t\to -\infty$.

The boundary condition (\ref{eq:boundary_conditions}) for $t\to -\infty$ allows for the fact that the system starts from the attractor and the optimal value of the noise is zero. A better understanding of this condition comes from the analysis of the equations of motion on the optimal trajectory.

\subsection{Optimal trajectory}
\label{subsec:optimal_trajectory}

It follows from Eqs.~(\ref{eq:var_problem}) that on the optimal trajectory
\begin{eqnarray}
\label{eq:var_eqns}
&&\frac{1}{2}\int\nolimits_{-\infty}^{\infty} dt' {\cal F}(t-t')f(t') = \chi(t),\\
&&\ddot\chi(t) +U_h''\bigl(q(t)\bigr)\chi(t) -\beta\int\nolimits_t^{\infty}dt'\phi_h(t' - t)\dot\chi(t')=0,\nonumber
\end{eqnarray}
while the interrelation between $f(t)$ and $q(t)$ is of the form (\ref{eq:eom}). Equation (\ref{eq:var_eqns}) for $\ddot\chi$  directly applies to the problem of escape discussed in Section~\ref{subsec:escape}. In the problem of reaching a given state, as seen from Eq.~(\ref{eq:boundary_conditions}), $\chi(t)$ is discontinuous at the instant $t=0$ when the system reaches the targeted state. Therefore $\dot\chi(t)$, along with a smooth part at $t<0$ and $t>0$, has a term $-\chi(-\ep)\delta(t-\ep)$ with $\ep\to +0$. 

If in the variational problem (\ref{eq:var_problem}) the integral over $t$ (the one with the integrand $\propto \chi(t)$) is taken from $-\infty$ to the instant of observation $t=0$, Eq.~(\ref{eq:var_eqns}) for $\ddot \chi$ reads
\begin{eqnarray}
\label{eq:chi_eqn}
\ddot\chi(t) &+&U_h''\bigl(q(t)\bigr)\chi(t) -\beta\int\nolimits_t^{0}dt'\phi_h(t' - t)\dot\chi(t')\nonumber\\
&&+\beta\phi_h(t)\chi(0)=0, \qquad t<0.
\end{eqnarray}
It coincides with Eq.~(\ref{eq:var_eqns}) for $\ddot\chi$ is one allows for the aforementioned $\delta$-function in $\dot \chi$.

From Eqs.~(\ref{eq:noise_distribution}) and (\ref{eq:var_eqns})
\begin{equation}
\label{eq:f_opt}
f(t)=\frac{1}{D}\int\nolimits_{-\infty}^\infty dt' \phi_f(t-t')\chi(t').
\end{equation}

An interesting and somewhat counterintuitive feature of Eqs.~(\ref{eq:var_eqns}) and (\ref{eq:chi_eqn}) is that the time evolution of $\chi(t)$ is acausal, the value of $\chi(t_1)$ depends on $\chi(t'_1)$ with $t'_1 > t_1$. This does not make the equations ill-defined, since we are solving a boundary-value problem, where we know where the system arrives at $t=0$ and what happens to the noise and $\chi(t)$ after that.

Equations (\ref{eq:eom}) and (\ref{eq:var_eqns}) are simplified near the attractor, where $U_h'(q)\approx \omega_a^2(q-q_a)$ and the equations become linear. In particular, for weak coupling to the bath $\chi(t)$ for $t \to -\infty$ has the form
\begin{equation}
\label{eq:chi_asympt}
\chi(t)=\chi_-\exp(-\lambda_{a-}t) +\chi_+\exp(-\lambda_{a+}t),
\end{equation}
where $\lambda_{a\pm}$ are given by Eqs.~(\ref{eq:stable_eigenvalues}) or (\ref{eq:overdamped_eigenvalue}) and $\chi_{\pm}$ are arbitrary constants. Since Re~$\lambda_{a\pm} <0$, the solution (\ref{eq:chi_asympt}) satisfies the boundary condition (\ref{eq:boundary_conditions}) for $t\to -\infty$. Generally, there are also other terms in $\chi(t)$, see below, but they decay faster than (\ref{eq:chi_asympt}) if the correlation time of the bath fluctuations is small compared to $|{\rm Re}~\lambda_{a\pm}|^{-1}$, as assumed in Eq.~(\ref{eq:chi_asympt}). 

By comparing Eqs.~(\ref{eq:var_eqns}) and (\ref{eq:eigenvalues}) one can see that, if $\lambda$ is a decrement that characterizes the approaching of the coordinate to the stable state in the absence of noise, then $-\lambda$ describes the increase of $\chi(t)$ near the stable state for $t\to -\infty$.

If the correlation time of the noise $f(t)$ is shorter than the relaxation time of the system, then asymptotically for $t \to -\infty$
\begin{eqnarray}
\label{eq:f_asympt}
&&f(t)=\sum\nolimits_{\nu = \pm}f_\nu \exp(-\lambda_{a\nu}t), \nonumber\\
&&f_\pm=\frac{\pi}{D}\chi_\pm \left[\Phi_f(i\lambda_{a\pm}) +\Phi_f(-i\lambda_{a\pm})\right].
\end{eqnarray}
Here, $\Phi_f(\omega)=\pi^{-1}\int\nolimits_0^\infty dt\phi_f(t)\exp(i\omega t)$; function $2{\rm Re}~\Phi_f(\omega)$ is the power spectral density of the noise $f(t)$. Equation (\ref{eq:f_asympt}) shows that on the optimal trajectory $f(t)\to 0$ for $t\to -\infty$, in agreement with the boundary condition (\ref{eq:boundary_conditions}).

The deviation of the coordinate and momentum of the system from $(q_a,p_a)$ is also exponential in time for $t\to -\infty$; from Eqs.~(\ref{eq:eom}) and (\ref{eq:f_asympt})  for a short noise correlation time
\begin{eqnarray}
\label{eq:q_asympt}
q(t)-q_a&&=-\sum\nolimits_{\nu=\pm}f_{\nu}\exp(-\lambda_{a\nu}t)\nonumber\\
&&\times\left\{\pi\beta\lambda_{a\nu}\left[\Phi_h(i\lambda_{a\nu}) + \Phi_h(-i\lambda_{a\nu})\right]\right\}^{-1}.
\end{eqnarray}

From Eqs.~(\ref{eq:chi_asympt}) - (\ref{eq:q_asympt}), the optimal trajectory is fully determined by the parameters $\chi_-$ and $\chi_+$. These parameters must be found from the boundary condition for $t=0$. 

\subsection{Escape problem}
\label{subsec:escape}

Noise can also lead to escape of the system from the initially occupied attractor and switching to another attractor. To find the probability of the corresponding large fluctuation per unit time one should again minimize the functional ${\cal R}_f$ with respect to noise realizations that lead to escape. The key here is to note that, after the noise $f(t)$ and the memory kernel $\phi_h(t)$ will have decayed, the system should be outside the basin of attraction of the initially occupied attractor or at least on the boundary of this basin. 

A correlated noise decays in time smoothly, it takes infinite time to decay to zero, as is also the case, generally, for the memory kernel.  On the other hand, for  $t\to \infty$ the system approaches a stationary state. From Eq.~(\ref{eq:chi_asympt}) such state may not be an attractor, since $\chi(t)$ would diverge there for $t\to\infty$. Therefore it must be a saddle point $\qS$, a local maximum of the potential $U_h(q)$. 

From the above arguments, which are similar to the corresponding arguments in systems without delay \cite{Dykman1990}, it follows that, to logarithmic accuracy,  the escape rate has the form 
\begin{equation}
W_e\propto \exp(-R_A/D),
\end{equation}
where the effective activation energy $R_A$ is given by the solution of the variational problem (\ref{eq:var_problem}) with the boundary condition for $t\to -\infty$ of the form of Eq.~(\ref{eq:boundary_conditions}), whereas the other boundary conditions are
\begin{equation}
\label{eq:boundary_escape}
f(t), \chi(t) \to 0,\quad q(t)\to \qS \quad {\rm for} \quad t\to \infty. 
\end{equation}

It follows from Eq.~(\ref{eq:var_eqns}) linearized near $\qS$ that, for weak coupling, $\chi(t)$ decays as $\exp(-\lambda_{{\cal S}+}t)$ for $t\to \infty$. If the correlation time of $f(t)$ is smaller than $1/\lambda_{{\cal S}+}$, on the optimal escape trajectory $f(t)$ and $q(t)-\qS$ decay in the same way. Otherwise the decay of $f(t)$ and $q(t)$ is controlled by the decay of $f(t)$ as given by Eq.~(\ref{eq:f_opt}) or the decay of $\phi_h(t)$. This completes the formulation of the problem of the activation energy of escape in the presence of delay.

We note that the explicit expression for $f(t)$ in terms of $\chi(t)$, Eq.~(\ref{eq:f_opt}), allows eliminating $f(t)$ and reducing the variational problem (\ref{eq:var_problem}) to that for two coupled functions, $\chi(t)$ and $q(t)$. In fact, it corresponds to integrating over realizations of $f(t)$, and the resulting functional (besides the $q$-dependent part) is the characteristic functional of the noise $\tilde{\cal P}_f[i\chi(t)]$ \cite{FeynmanQM}. The formulation can be extended also to non-Gaussian noise, and there, too, one can use the characteristic functional of the noise; the boundary conditions for the optimal fluctuation to a given state of the system and for escape are the same as for Gaussian noise. 

\section{Large fluctuations induced by thermal noise}
\label{sec:thermal_noise}

Optimal trajectories of the system and the noise can be found and the logarithm of the probability distribution can be obtained in an explicit form  in the case where the noise $f(t)$ is of purely thermal origin, $f(t)=h_{\rm b}^{(0)}(t)$. In this case $\phi_f(t) = \phi_h(t)$ and the noise intensity $D=k_BT\equiv 1/\beta$. One can show that the variational equations for $q(t), f(t)$, and $\chi(t)$,  Eqns.~(\ref{eq:eom}), (\ref{eq:var_eqns}) and (\ref{eq:chi_eqn}), for the time prior to reaching the ``target" point $(q,p)$ at $t = 0$ have a solution:
\begin{eqnarray}
\label{eq:thermal_traj}
\chi(t) &=&\dot q(t), \qquad t\leq 0,\nonumber\\
 \ddot q(t)+U_h'\bigl(q(t)\bigr)&=& \beta\int_t^0dt'\phi_h(t' -t)\dot q(t');
\end{eqnarray}
function $f(t)$ is given by Eq.~(\ref{eq:f_opt}) with $\chi(t)=\dot q(t)$ for $t\leq0$ and $\chi(t)=0$ for $t>0$.

The solution (\ref{eq:thermal_traj}) applies also to the problem of escape, except that the integral in the right-hand side of the second equation goes from $t$ to $\infty$ and the solution applies in the whole range $-\infty< t < \infty$. 

Noting that in the variational problem (\ref{eq:var_problem}) ${\cal R}_f[f(t)]=\frac{1}{2}\int_{-\infty}^\infty dt \chi(t) f(t)$ and expressing $f(t)$ in this expression in terms of $q$ from the equation of motion of the system (\ref{eq:eom}), one obtains
\begin{eqnarray}
\label{eq:Boltzmann}
R(q,p) &=& \frac{1}{2}p^2 + U_h(q)-U_h(q_a), \nonumber\\
R_A &=& U_h(\qS)- U_h(q_a).
\end{eqnarray}
Thus, as expected, in thermal equilibrium  the probability distribution of the system (\ref{eq:prob_density}) is of the Boltzmann form, with a renormalized potential due to the interaction (however, our calculation does not give the prefactor of the distribution). The activation energy of escape is given by the renormalized height of the potential barrier.

We now discuss the form of the optimal trajectories (\ref{eq:thermal_traj}) for $\delta$-correlated fluctuations of the  thermal bath, where $\phi_h(t)=4\Gamma k_BT\delta (t)$, with $\Gamma$ being the coefficient of viscous friction. In this case the trajectory $q(t)$, Eq.~(\ref{eq:thermal_traj}), differs from the trajectory $q(t)$ in the absence of noise, Eq.~(\ref{eq:eom}) with $f(t)=0$, by time inversion, $t\to -t, \dot q \to -\dot q$, cf.~\cite{Onsager1953}. Using the ``prehistory problem" formulation \cite{Dykman1992d}, this symmetry was seen in simulations and experiments. As mentioned in the Introduction, it can be understood by noting that in a fluctuation the system gets energy at the expense of the decrease of entropy of the thermal bath. Taking a time-reversed path minimizes the required entropy change. 

Care must be taken when applying the time-reversal argument in the general case of delayed dissipation. 
Indeed, to define a noise-free trajectory it is insufficient to specify the starting point in phase space $(q,p)$; instead one has to specify the whole history of motion before the state $(q,p)$ was reached. 

\subsection{Symmetry of the most probable trajectories}
\label{subsection:symmetry_breaking}


The previous analysis referred to the comparison of the most probable trajectories that come to a given state as a result of a fluctuation and the trajectories that start at this state and in the absence of noise go to the attractor. Instead one can consider trajectories that go through a given state. They reach it from the vicinity of the attractor as a result of a fluctuations, and then, as the fluctuation decays, they go back to the attractor. The distribution of such trajectories was studied for Markovian systems by Luchinsky and McClintock \cite{Luchinsky1997b}, It was shown that, for an overdamped system in thermal equilibrium, the distribution peaks at a trajectory that is symmetric with respect to time reversal, if time is counted off from the instant of reaching a chosen state. In other words, the segments of the most probable trajectory, which correspond approaching the state and moving away from it, are symmetrical. 

We now consider the most probable trajectory that goes through a given state $(q,p)$ for systems with delayed dissipation. We assume that the trajectory passes this state at $t=0$ (obviously, this instant is arbitrary). The segment of the trajectory for $t<0$ is the optimal fluctuational trajectory. In thermal equilibrium it is described by Eq.~(\ref{eq:thermal_traj}). The most probable trajectory after the state $(q,p)$ has been reached is described by Eq.~(\ref{eq:eom}). In contrast to Markov systems, the force $f(t)$ in this equation is generally nonzero, because the noise is correlated. Since it was nonzero for $t<0$, where it drove the system against the potential gradient, it does not instantly go to zero for $t>0$. The most probable value of $f(t)$ for $t>0$ is given by Eq.~(\ref{eq:f_opt}) with $D=k_BT$. This value maximizes the probability density of a realization of $f(t)$ that brings the system to state $(q,p)$. 

From Eqs.~(\ref{eq:eom}) and (\ref{eq:f_opt}) with account taken of the relations $\chi(t)=\dot q$ for $t<0$ and $\chi(t)=0$ for $t>0$, we obtain that the most probable trajectory of the system after the state $(q,p)$ has been reached is described by equation
\begin{equation}
\label{eq:traj_after}
 \ddot q(t)+U_h'\bigl(q(t)\bigr)= - \beta\int_0^tdt'\phi_h(t -t')\dot q(t') \quad (t>0).
\end{equation}

A comparison of Eqs.~(\ref{eq:thermal_traj}) and (\ref{eq:traj_after}) shows that, in thermal equilibrium, for the most probable trajectory that goes through a state $(q,p)$ with $p=0$, the segments that lead to the state from the attractor [Eq.~(\ref{eq:thermal_traj})] and from the state to the attractor [Eq.~(\ref{eq:traj_after})] have time-reversal symmetry $t\to -t, \dot q \to -\dot q$. This symmetry is illustrated in Fig.~\ref{fig:trajectories} below.

We emphasize again that, in contrast to Markov systems, one cannot compare separately the most probable trajectory to a given state and the most probable trajectory from a given state, because the latter is not defined. Only a single optimal trajectory that goes through a given state has the symmetry. This trajectory is fully determined by the state: the integrals in Eqs.~(\ref{eq:thermal_traj}) and (\ref{eq:traj_after}) go to/from $t=0$, where the system is at the given $(q,p)$. Another distinction from Markov system is that the trajectory is smooth at $(q,p)$ even where $p\neq 0$.

\section{Thermal noise with exponentially correlated power spectrum}
\label{sec:expon_correlated_noise}

To illustrate the general results we will discuss the case where the noise spectrum has a simple form,
\begin{equation}
\label{eq:expon_corr}
\phi_h(t)=C_h\ka k_BT\exp(-\ka|t|),
\end{equation}
where $\ka = 1/t_c$ is the reciprocal correlation time and $C_h$ characterizes the noise intensity; for the coupling model (\ref{eq:Feynman_vernon}) $C$ is independent of temperature. The limit $\ka \to \infty$ corresponds to $\delta$-correlated noise. 

For the model (\ref{eq:expon_corr}),  Eq.~(\ref{eq:eigenvalues})  for the eigenvalues that characterize noise-free motion near attractor takes the form
\begin{equation}
\label{eq:eigenvalues_expon}
(\lambda + \ka)(\lambda^2+\omega_a^2) + C_h\ka\lambda =0.
\end{equation}
If the coupling is weak, two of the roots of this equation $\lambda_{1,2}$ are given by Eqs.~(\ref{eq:stable_eigenvalues}) for $C_h\ka^2/(\ka^2+\omega_a^2)\ll \omega_a$ or Eq.~(\ref{eq:overdamped_eigenvalue}) for $\ka\gg C_h\gg \omega_a$, respectively. In addition, Eq.~(\ref{eq:eigenvalues_expon}) has a root with a much larger (in the absolute value) real part 
\begin{equation*}
\lambda_3\approx -\ka +C_h\ka^2/(\ka^2+\omega_a^2), \qquad C_h\ll \ka
\end{equation*}
($-\lambda_3\gg -{\rm Re}~\lambda_{a\pm}$). 

In the opposite limit $C_h\gg \ka \gg \omega_a$, which can be of relevance for the model (\ref{eq:Feynman_vernon}),  the roots become $\lambda_{1}\approx - \omega_a^2/C_h$ and $\lambda_{2,3}\approx -\left[\ka \mp i\sqrt{4\ka C_h-\ka^2}\right]/2$. For $C_h\sim \ka \sim \omega_a$ the real parts of all 3 roots $\lambda_{1,2,3}$ are of the same order of magnitude. It is easy to see that $|{\rm Re}~\lambda_{1,2,3}| < \ka$, i.e., correlations of the bath decay faster than $q(t)$. 

An explicit solution of Eq.~(\ref{eq:thermal_traj}) for the optimal trajectory to a state $(q,p)$ can be obtained for a harmonic potential $U_h(q)=\omega_a^2q^2/2$. It reads 
\begin{eqnarray}
\label{eq:optimal_parabolic}
&&q(t)=\sum_{i}q_i e^{-\lambda_it},\qquad \sum_{i}\lambda_i(\ka +\lambda_i)^{-1}q_i =0,\nonumber\\
&&\sum_{i}q_i=q,\qquad \sum_{i}\lambda_iq_i=-p \qquad (i=1,2,3).
\end{eqnarray}
For the coupling to the bath of the form (\ref{eq:Feynman_vernon}), this solution applies for an arbitrary coupling strength. It shows that the optimal fluctuational trajectory is a superposition of three exponentials.

\begin{figure}[h]
\centering
\includegraphics[scale=0.3]{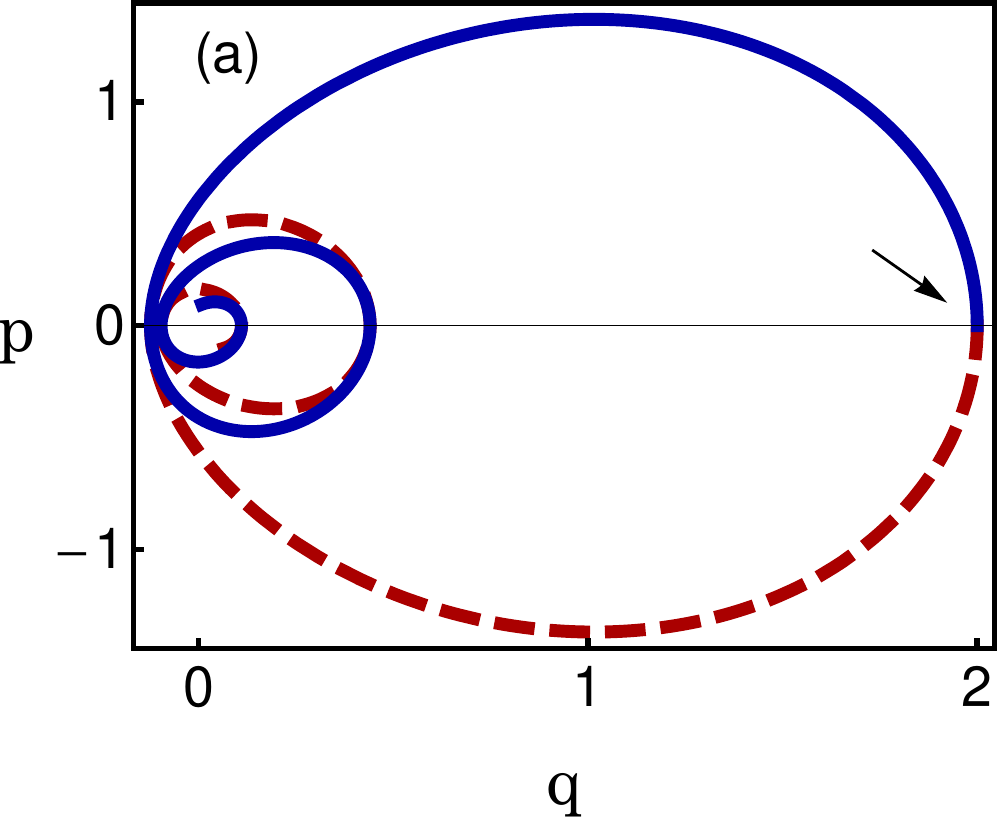}
\hspace{0.2in} \includegraphics[scale=0.315]{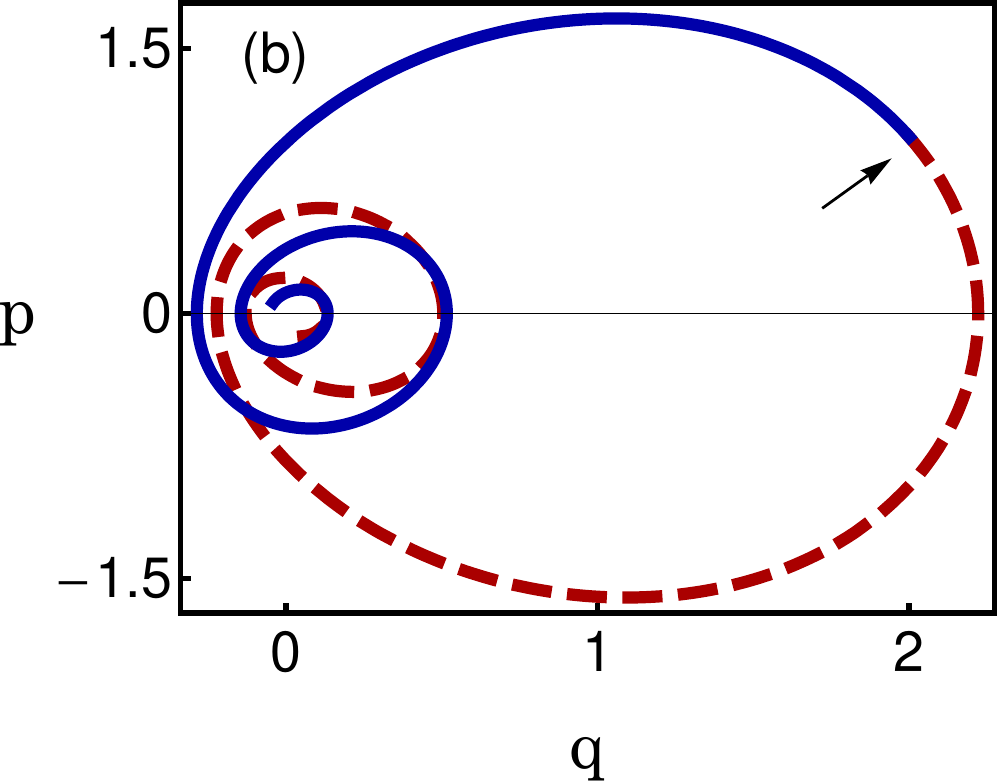}\\
\hfill\\
\includegraphics[scale=0.3]{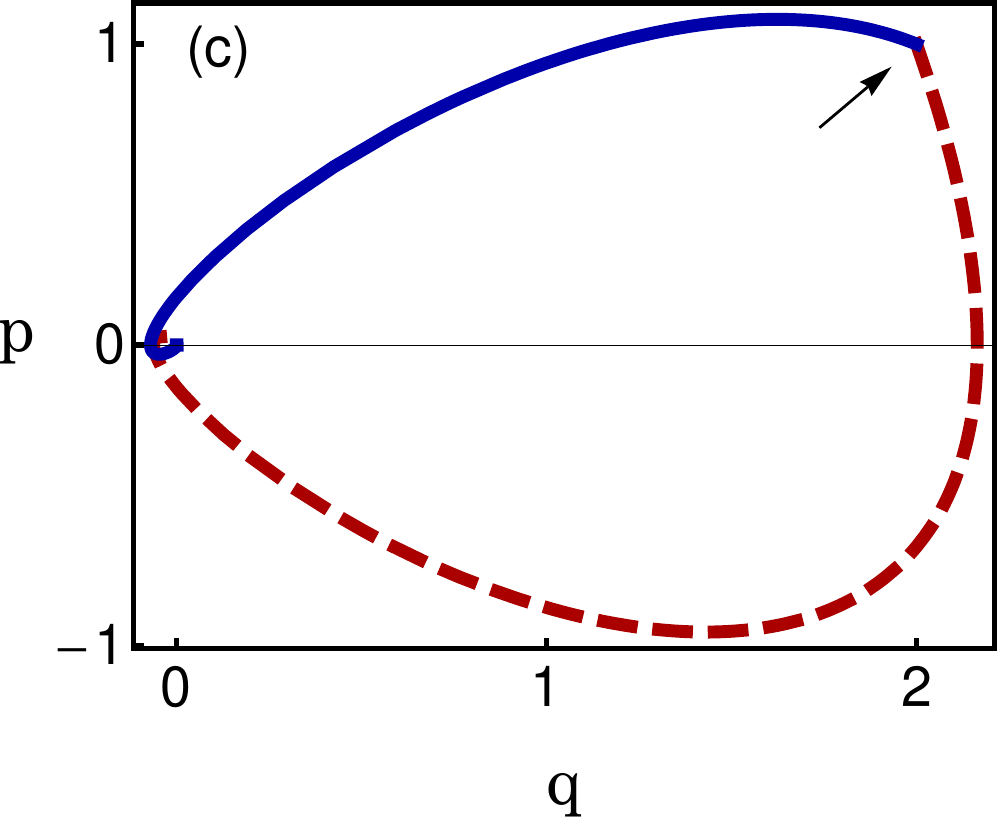}
\hspace{0.2in} \includegraphics[scale=0.3]{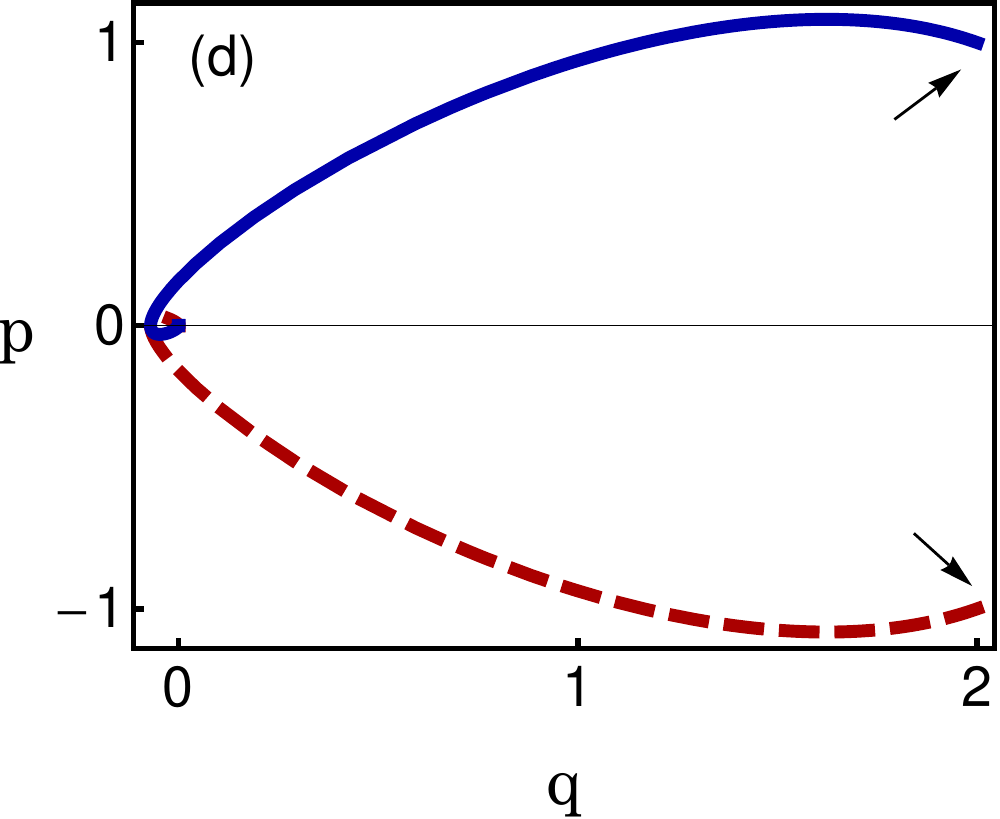}
\caption{The most probable trajectories of thermal fluctuations in which the system reaches targeted states (q,p) (the solid lines). The dashed lines show the sections of the trajectories after the targeted states have been reached. Plots (a) and (b) refer to a system in a parabolic potential, $U_r=\omega_a^2q^2/2$, and the exponentially correlated noise, Eq.~(\protect\ref{eq:expon_corr}), with $\ka/\omega_a=1, C_h/\omega_a=1.5$. The targeted states are $(q=2,p=0)$ in (a) and $(q=2,p/\omega_a=1)$ in (b), as indicated by the arrows. Panel (c) shows the same trajectories as in (b) in the Markov limit $\ka \to \infty$. Panel (d) shows the symmetry of the most probable trajectories to a state $(q,p)$ and from the state $(q,-p)$ in Markov systems in thermal equilibrium.}
\label{fig:trajectories}
\end{figure}

The most probable trajectories for reaching a given state $(q,p)$ and then moving back to the attractor for the model (\ref{eq:expon_corr}) are shown in Fig.~\ref{fig:trajectories}. The sections of the trajectories to and from the targeted state are symmetric for $p=0$. The symmetry is lost if $p\neq 0$, but the overall trajectory to and from the state is smooth. In contrast, for Markov systems,  on the most probable trajectory that reaches a state $(q,p)$ the derivative $dp/dq$ is discontinuous at this state for $p\neq 0$. However, in such systems the most probable trajectory to state $(q,p)$ has a symmetric noise-free counterpart that starts from $(q,-p)$ and goes to the attractor. Such counterpart is not generally defined for systems with delay.

For a nonparabolic potential $U(q)$ one can find the trajectory to a given state $(q,p)$ numerically by integrating Eq.~(\ref{eq:thermal_traj}) backward in time from $t=0$, using the values of $q(0)=q$ and $\dot q(0)=p$. To find the optimal escape trajectory, on the other hand, one can differentiate Eq.~(\ref{eq:thermal_traj}) over time, which leads to equation 
\begin{equation}
\label{eq:third_order}
\dddot q + \ka \ddot q + [U_h''(q)+C_h\ka]\dot q -\ka U_h'(q)=0
\end{equation} 
One can then seek the solution of this equation by the shooting method, starting from $t\to -\infty$, where it has the form of a sum of exponentials, see Eq.~(\ref{eq:optimal_parabolic}). The coefficients $q_{1,2,3}$ have to be found from the condition that the trajectory arrives at $(\qS,p=0)$ for $t\to \infty$ and that the solution satisfies the initial intego-differential equation (\ref{eq:thermal_traj}), which leads to the relation between $q_{1,2,3}$ given by the second equation in (\ref{eq:optimal_parabolic}).

\section{Short noise correlation time}
\label{sec:short_corr_time}

An explicit solution for the probability distribution and the escape rate can be obtained in the case of short correlation time of the noise and short delay time of the bath, when functions $\phi_f(t)$ and $\phi_h(t)$ are close to $\delta$-functions. To do this it is convenient to eliminate $f(t)$ from the functional (\ref{eq:var_problem}). Then the variational problem for reaching a given state $(q,p)$ takes the form
\begin{eqnarray}
\label{eq:var_probl_chi}
R(q,p)=&&\min\left\{ -\frac{1}{D}\int_{-\infty}^0dt\int_{-\infty}^t dt'\chi(t)\phi_f(t-t')\chi(t')
\right.
\nonumber\\
&&\left.
 + \int_{-\infty}^{0}dt\,\chi(t)\Bigl[\ddot q(t) + U_h'\bigl(q(t)\bigr)
\Bigr.\right.\nonumber\\
&&\left.\left.
+\beta\int_{-\infty}^tdt' \phi_h(t-t')\dot q(t')\right]\right\}.
\end{eqnarray}
The variational problem for the activation energy of escape is given by Eq.~(\ref{eq:var_probl_chi}) in which the integrals over $t$ go from $-\infty$ to $\infty$ rather than from $-\infty$ to $0$.

If the dissipation has no delay and the noise is white, $\phi_f(t)=4\Gamma D\delta(t)$ and $\phi_h(t)=4\Gamma k_BT\delta(t)$, the optimal trajectories for the variational problem (\ref{eq:var_probl_chi}) are
\begin{equation}
\label{eq:opt_traj_no_delay}
\ddot q +U_h' -2\Gamma\dot q=0,\qquad \chi=\dot q.
\end{equation}
In this approximation, which is of the zeroth-order in delay and correlation, 
\begin{eqnarray}
\label{eq:zeroth_order}
&&R^{(0)}(q,p)=\frac{1}{2}p ^2+ U_h(q)-U_h(q_a),\nonumber\\
&&R^{(0)}_A=U_h(\qS) - U_h(q_a).
\end{eqnarray}

If the correlation time of $\phi_f,\phi_h$ is short compared to the relaxation time of the system, one can write 
\begin{equation*}
\phi_{f,h}(t)=2\pi\Phi_{f,h}(0)\delta(t) + \left[\phi_{f,h}(t)
-2\pi\Phi_{f,h}(0)\delta(t)\right]
\end{equation*}
and then consider the term in the brackets as a perturbation. We choose $\Gamma$ and $D$ in the same way as for $\delta$-correlated  $\phi_{f,h}(t)$, 
\begin{eqnarray*}
2\pi\Phi_f(0)&\equiv& \int_{-\infty}^\infty dt \phi_f(t)=4\Gamma D,\\
2\pi\Phi_h(0)&\equiv& \int_{-\infty}^\infty dt \phi_h(t)=4\Gamma k_BT.
\end{eqnarray*}
To the first order in the delay/correlation perturbation, the correction to $R(q,p)$ can be calculated using the zeroth order trajectories (\ref{eq:opt_traj_no_delay}). In the integrals over $t'$ in Eq.~(\ref{eq:var_probl_chi}) it is convenient to expand $\chi(t')\approx \chi(t) + (t'-t)\dot\chi(t)$ and $\dot q(t')\approx \dot q(t) +(t'-t)\ddot q(t)$. Substituting this into Eq.~(\ref{eq:var_probl_chi}) one obtains
\begin{eqnarray}
\label{eq:correction_distribution}
&&R(q,p)\approx R^{(0)}(q,p) + \Gamma (\overline {t_f} - \overline {t_h})p^2,\nonumber\\
&&\overline{t_{f,h}}=\int_0^{\infty}t\phi_{f,h}(t)dt\Big/\int_0^{\infty}\phi_{f,h}(t)dt.
\end{eqnarray}

The parameters $\overline{t_{f}}$ and $\overline{t_h}$ characterize the widths of the correlators $\phi_{f}(t)$ and $\phi_h(t)$, respectively. From Eq.~(\ref{eq:correction_distribution}), the correction to $R(q,p)$ is generally of the first order in these widths (cf. Ref.~\onlinecite{Dykman1990} where there was found a correction due to noise correlations for an overdamped system with no delay). The broadening of the noise correlator $\phi_f(t)$ and the dissipation delay [the broadening of $\phi_h(t)$] act in opposite directions: the larger is $\overline{t_f}$ the smaller is the probability (\ref{eq:prob_density}) to reach a remote state, whereas the increase of $\overline{t_h}$ increases this probability. In thermal equilibrium the two effects compensate each other.

From Eq.~(\ref{eq:correction_distribution}), there is no correction to the activation energy of escape $R_A$ of the first order in the width of the correlators $\phi_{f,h}(t)$, because $p=0$ at the saddle point. The lowest-order correction appears in the second order. It can be calculated similarly to the above procedure, taking into account that the integrals over $t$ in Eq.~(\ref{eq:var_probl_chi}) now run to $\infty$. In the integrals over $t'$ one should expand  $\chi(t')$ and $\dot q(t')$ about $\chi(t)$ and $\dot q(t)$, respectively, to the second order in $t-t'$. Then Eq.~(\ref{eq:opt_traj_no_delay}) gives
\begin{eqnarray}
\label{eq:correction_activation}
&&R_A\approx R_A^{(0)}  + \Gamma\left(\overline {t_f^2} - \overline {t_h^2}\right)\int_{-\infty}^{\infty}dt \ddot q^2(t),\nonumber\\
&&\overline{t_{f,h}^2}=\int_0^{\infty}t^2\phi_{f,h}(t)dt\Big/\int_0^{\infty}\phi_{f,h}(t)dt,
\end{eqnarray} 
where $\ddot q(t)$ is given by Eq.~(\ref{eq:opt_traj_no_delay}) with boundary conditions $q(t)\to q_a$ for $t\to -\infty$ and $q(t)\to \qS$ for $t\to \infty$. Again, the changes of the escape activation energy $R_A$ due noise correlations and dissipation delay have opposite signs and compensate each other for thermal noise.

\section{Conclusions}
\label{sec:conclusions}

In this paper we have considered large rare fluctuations induced by Gaussian noise in systems with delayed dissipation. The dissipation comes from coupling to a thermal bath; the corresponding friction force depends on the history of the system motion and is described by an integral of the velocity with an appropriate kernel. The noise, along with the part that comes from the thermal bath, can have another source. 

The proposed formulation reduces the problems of finding the logarithm of the probability distribution over the phase space of the system and the effective activation energy of escape from a metastable state to variational problems. The extreme trajectories of the respective variational functionals provide the most probable paths followed by the system in a fluctuation to a given state or in escape and also the most probable corresponding realizations of the noise.

We show that, if the noise is coming from the thermal bath responsible for the dissipation, the logarithm of the probability distribution has a familiar Boltzmann form, with a renormalized potential due to the coupling to the bath. Closed-form equations for the most probable system trajectories are obtained in this case. We describe both the portion of the trajectory leading to a state and followed after the state had been reached. It is shown that, if the state corresponds to zero momentum, these portions are time-reversal symmetric. In contrast to Markov systems, however, the phase trajectories are smooth at the observation point. This is a clear signature of the dissipation delay. Also, in contrast to Markov systems, one may not compare the most probable trajectory to a given point $(q,p)$ in the phase space of the system with the trajectory from this point, since the latter is not defined for a system with delayed dissipation. Explicit solutions are obtained for the exponentially correlated in time noise and are used to illustrate the general properties of the trajectories.

The case of the noise and dissipation with short correlation times is analyzed. It is shown that the logarithm of the probability distribution has corrections of the first order in the widths of the time correlation function of the noise and the dissipation kernel.  In contrast, the effective activation energy of switching has only second-order corrections in these widths.

Delayed dissipation is commonly encountered in physical systems, the simplest example being dissipation from the coupling to an electromagnetic environment with frequency-dependent impedance. This makes it possible to test the predictions of the present paper in experiments similar to those used to observe optimal fluctuational paths in effectively Markovian systems, in particular in nano- and micromechanical resonators. Underdamped resonators with strong nonlinearity would be advantageous for such experiments, as there the vibration frequency strongly varies as a result of energy fluctuations. Therefore the density of states of the excitations of the thermal bath that resonate with the system and cause energy absorption varies as well. Another candidate systems of significant current interest are optomechanical systems in high-Q optical cavities, where the density of states of the electromagnetic bath has strong  dispersion.

The research of M.I.D. was supported in part by  NSF Grant No. CMMI-0900666 and DARPA through the DEFYS program.


%

 \end{document}